\documentclass[conference,letterpaper]{IEEEtran}

\usepackage[top=0.75in, bottom=1.05in, left=0.65in, right=0.65in]{geometry}

\usepackage[utf8]{inputenc}
\usepackage[T1]{fontenc} 
\usepackage{url}
\usepackage{ifthen}
\usepackage{cite}
\usepackage[cmex10]{amsmath}
\usepackage{graphicx}
\usepackage{booktabs}
\usepackage{amssymb}
\usepackage{algorithm}
\usepackage{algpseudocode}

\newtheorem{thm}{Theorem}
\newtheorem{defn}[thm]{Definition}

\begin{document}
\title{Automated Proving of Shannon-Type Entropy Inequalities via Fine-Tuned Language Models and Guided Tree Search}

\author{%
\IEEEauthorblockN{
Shing Yin Wong$^{\dagger}$,
Shaocheng Liu$^{*}$,
Linqi Song$^{\dagger}$,
Amin Gohari$^{*}$,
Cheuk Ting Li$^{*}$
}
\IEEEauthorblockA{
$^{\dagger}$Department of Computer Science, City University of Hong Kong\\
$^{*}$Department of Information Engineering, The Chinese University of Hong Kong\\
}
}


\maketitle

\begin{abstract}
THIS PAPER IS ELIGIBLE FOR THE STUDENT PAPER AWARD.
Proving Shannon-type entropy inequalities is a fundamental task in information theory that often requires constructing non-trivial linear combinations of known constraints, which is a combinatorial search problem that scales poorly with the number of random variables.
We investigate whether small-scale large language models (0.6B--1.7B parameters), fine-tuned on atomic proof steps and combined with guided beam search, can automate this process.
On a held-out test set of 60 inequalities spanning n=10 to 15 variables, our 0.6B fine-tuned model achieves an 85\% proof success rate with tree search. GPT-5.5 solves 1.7\% samples under zero-shot prompting while Psitip solves 33.3\% samples.
A systematic ablation study across training context length (4096 vs.\ 8192 tokens) and data distribution (n=9-skewed vs not skewed) reveals that a 4096-token not skewed training distribution yields the best performance, with extended context and skewed data providing no marginal benefit.
We further identify two dominant failure modes---format failures and step quality degradation---and verify that the beam-scoring heuristic is essential via a controlled ablation (random scoring reduces success from 83\% to 23\%).
\end{abstract}

\section{Introduction}

Entropy inequalities are fundamental tools in information theory. They are widely used to give outer bounds in specific problems such as network coding~\cite{yeung1998}, secret sharing~\cite{secretsharing}, regenerating codes~\cite{RegeneratingCodes}, and related problems. 

Given a target inequality, determining whether it is a valid entropy inequality is quite difficult, especially when the number of random variables is large.
To address this problem, Yeung introduced a linear programming framework for proving information inequalities~\cite{itip}, which was implemented in the MATLAB package Information Theoretic Inequality Prover (ITIP). It verifies a target inequality by checking whether it can be written as a non-negative linear combination of Shannon-type inequalities and additional conditions. Thus, the proof is reduced to a linear programming problem. Several automated tools were later developed along this line. Xitip~\cite{xitip} improves the implementation by using a C-based solver, while Minitip~\cite{minitip} provides a lightweight C-based prover with a simple syntax. Aitip~\cite{Aitip} offers an online service for proving or disproving information inequalities. Psitip~\cite{psitip} further extends automated reasoning to a symbolic Python environment, supporting constrained, auxiliary-variable, existential, and non-Shannon-type reasoning. Recently Gaussian-elimination methods~\cite{gauss} aim to reduce the size of the LP problem and make the proof process more efficient.
 
However, for an inequality involving $n$ random variables, the space of possible proof-step sequences is vast. Since the LP search space grows exponentially, LP-based methods face memory and compute limitations as $n$ grows beyond $10$.

Recent advances in large language models (LLMs) have shown promise in formal mathematical reasoning, including theorem proving and proof automation~\cite{gptf,polu2022formal,kaliszyk2018,wang2017premise,htps,thor,deepseekprover,bfsprover,proofoptimizer,prod,palm,leancopilot,processautoformalization} and mathematical/olympiad problem solving~\cite{deepseekmath,alphageometry,alphageometry2,alphaproof,aips,frontiermath}.
However, the application of LLMs to \emph{entropy inequality proving} remains largely unexplored.
Entropy inequalities present a unique challenge: they require both the semantic intuition to identify which variables to condition on, and the syntactic precision to produce machine-verifiable proof steps with correct indices and coefficients---the latter being precisely where frontier LLMs struggle.

In this work, we investigate whether small-scale, supervised fine-tuned (SFT) large language models, combined with a guided beam-search tree search, can automate the proof of Shannon-type entropy inequalities.
Our key contributions are:

\begin{enumerate}
  \item We demonstrate that a 0.6B-parameter SFT model with tree search achieves 85\% proof success on a challenging test set of 60 inequalities ($n=10$--$15$). Frontier LLMs (GPT-5.5, Claude-4.6 Sonnet) score only 1.7\% (1/60) under zero-shot prompting.
  \item Through a systematic four-dataset ablation (A/B/C/D), we show that a 4096-token not skewed training distribution is sufficient and optimal; extending context to 8192 tokens or skewing toward higher $n$ provides no benefit and can degrade robustness.
  \item We identify and characterize two primary failure modes---format failures and step quality degradation---and quantify the essential role of the beam-scoring heuristic through controlled ablation.
  \item We show advantages of our method for large $n$ compare to LP-solvers.
\end{enumerate}

\section{Problem Formulation}

\subsection{Shannon-Type Entropy Inequalities}

We review the basic concepts of Shannon-type entropy inequalities \cite{itip}.
Let $X_1, X_2, \ldots, X_n$ be discrete random variables. For a subset $S \subseteq [n] := \{1,\ldots,n\}$, the joint entropy of the variables $X_i$ where $i\in S$ is denoted as $H(X_{S})$.
\begin{defn}
A \emph{single entropy inequality} on the random variables $X_{1},\ldots,X_{n}$ is a sequence $(\alpha_{i},S_{i})_{i\in[k]}$ where $\alpha_{i}\in\mathbb{Z}$ and $S_{i}\subseteq[n]$, which represents the inequality
\[
\sum_{i=1}^{k}\alpha_{i}H(X_{S_{i}})\ge0.
\]
A \emph{conditional entropy inequality} is given by a single entropy inequality $(\alpha_{i},S_{i})_{i\in[k]}$ (the \emph{consequent}) together with a list of entropy inequalities $((\beta_{j,i},T_{j,i})_{i\in[\ell_{j}]})_{j\in[m]}$ (the \emph{conditions}), which represents the implication
\[
\Big(\forall j\in[m]:\,\sum_{i=1}^{\ell_{k}}\beta_{j,i}H(X_{T_{j,i}})\ge0\Big)\;\Rightarrow\;\sum_{i=1}^{k}\alpha_{i}H(X_{S_{i}})\ge0.
\]
\end{defn}
\medskip{}

\begin{defn}
A conditional entropy inequality is \emph{Shannon-type} if it is implied by the polymatroid inequality \cite{itip} $I(X_A; X_B | X_C) \ge 0$, or equivalently,
\[
H(X_{A\cup C})+H(X_{B\cup C})\ge H(X_{A\cup B\cup C})+H(X_{C}),
\]
for every $A,B,C\subseteq[n]$. More precisely, it is Shannon-type if
\begin{align*}
\sum_{i=1}^{k}\alpha_{i}H(X_{S_{i}})
&\equiv
\sum_{j=1}^{K}\mu_{j}
\bigl(
H(X_{A_{j}\cup C_{j}})
+H(X_{B_{j}\cup C_{j}}) \\
&\quad
-H(X_{A_{j}\cup B_{j}\cup C_{j}})
-H(X_{C_{j}})
\bigr) \\
&\quad
+\sum_{j=1}^{m} \nu_j
\sum_{i=1}^{\ell_{k}}
\beta_{j,i}H(X_{T_{j,i}}).
\end{align*}
for some $K\in\mathbb{Z}_{\ge0}$, $\mu_{j}\in\mathbb{Q}_{\ge0}$, $A_{j},B_{j},C_{j}\subseteq[n]$ for $j\in[K]$, and $\nu_{j}\in\mathbb{Q}_{\ge0}$ for $j\in[m]$. The datum $((\mu_{j},A_{j},B_{j},C_{j})_{j\in[K]},(\nu_{j})_{j\in[m]})$ is called a \emph{proof} of the conditional entropy inequality with \emph{length} $K$.
\end{defn}

The proof task is to certify the target inequality under the given conditions by
finding coefficients $(\mu_j, A_j, B_j, C_j)_{j\in[K]}$ and $(\nu_j)_{j\in[m]}$ certifying the decomposition.

\subsection{Proof via Stepwise Residual Reduction}

Our proof strategy follows an iterative decomposition: starting with residual $R_0 = I_{\text{target}}$, at each step $t$, we propose an atomic non-negative term $T_t \ge 0$. Where $T_t$ is either  a single information measure with positive coefficient (e.g., $H(A)$, $H(A|B)$, $I(A;B)$, and $I(A;B|C)$), or a positive scalar multiple of a given condition $\sum_{i=1}^{\ell_{k}}
\beta_{j,i}H(X_{T_{j,i}})$. Then we subtract it from the current residual:
\begin{equation}
  R_{t+1} = R_t - T_t.
\end{equation}
The proof succeeds if we reach $R_t = 0$.
This constraint ensures every step is automatically verifiable by the symbolic checker without requiring an LP solver at each step.

\subsection{Proof Example}

To illustrate the task, consider a concrete $n=14$ inequality from the test set.
The target is a 22-term inequality:
{\footnotesize
\begin{equation*}
\begin{split}
I_{\text{target}} = &\; -H(X_{1,2,5,6,9,11}) + H(X_{1,2,4,5,6,7,8,9,11}) \\
  & - 3H(X_{1,4,5,6,8,13}) - 3H(X_{1,6,8,11,13}) \\
  & + 3H(X_{1,6,7,8,11,12,13}) - H(X_{3,5,7,9,11,14}) \\
  & + H(X_{1,2,5,6,7,9,11,14}) - H(X_{1,2,4,5,6,7,8,9,11,14}) \\
  & - 3H(X_{1,3,4,5,6,8,12,14}) + 3H(X_{1,3,4,5,6,7,8,9,10,12,14}) \\
  & - 6H(X_{2,4,5,7,13,14}) + 6H(X_{2,4,5,6,7,8,9,13,14}) \\
  & + H(X_{2,3,5,6,7,8,9,11,13,14}) + 3H(X_{1,4,5,6,7,8,10,11,13,14}) \\
  & + 3H(X_{1,5,6,7,8,9,10,11,13,14}) + H(X_{3,5,6,7,8,9,10,11,13,14}) \\
  & - H(X_{2,3,5,6,7,8,9,10,11,13,14}) \\
  & - H(X_{1,2,4,5,6,7,8,9,10,11,13,14}) \\
  & + 3H(X_{1,3,4,5,6,7,8,12,13,14}) \\
  & - 3H(X_{1,5,6,7,8,9,10,11,12,13,14}) \\
  & + H(X_{1,2,4,5,6,7,8,9,10,11,12,13,14}) \\
  & - 3H(X_{1,3,4,5,6,7,8,9,10,11,12,13,14}) \ge 0.
\end{split}
\end{equation*}}
Our 0.6B model under beam search (beam 32, rollouts 6, depth 24) finds the following 7-step proof:
\begin{enumerate}
  \item $6H(X_{6,8,9}|X_{2,4,5,7,13,14}) \ge 0$
  \item $3I(X_{7,10,11,14}; X_{3,7,12,13,14}|X_{1,4,5,6,8,13}) \ge 0$
  \item $3I(X_{7,9,10}; X_{7,10,11,13}|X_{1,3,4,5,6,8,12,14}) \ge 0$
  \item $3I(X_{7,12}; X_{5,7,9,10,14}|X_{1,6,8,11,13}) \ge 0$
  \item $I(X_{4,7,8}; X_{7,14}|X_{1,2,5,6,9,11}) \ge 0$
  \item $I(X_{2,6,8,13}; X_{6,8,10,13}|X_{3,5,7,9,11,14}) \ge 0$
  \item $H(X_{12}|X_{1,2,4,5,6,7,8,9,10,11,13,14}) \ge 0$.
\end{enumerate}
Each step is independently verifiable as a Shannon-type entropy inequality. The residual after step~7 is exactly zero.

\section{Method}

\subsection{Training Data Generation}

We generate supervised training data using a reverse proof construction procedure.
Specifically, we randomly sample Shannon-type entropy inequalities and conditions, and combine them to construct a target inequality with a guaranteed
proof. Since our training framework is step-wise, each prompt corresponds to the
current residual inequality, and the model is expected to output only one proof
step at a time. Therefore, we need a principled strategy to select the next proof
step from the candidate Shannon inequalities and conditions.

Given a current residual inequality, we focus on entropy terms with negative
coefficients, since these terms indicate the part that still needs to be
eliminated in the remaining proof. For each candidate proof atom, we define a
score on the negative part magnitude of the current residual.
The candidate with the largest score is selected as the next output step.
See Algorithm \ref{alg:mix-type-generation}.
\begin{algorithm}[htbp]
\caption{Mixed-type training sample generation}
\label{alg:mix-type-generation}
\begin{algorithmic}[1]
\State Sample the number of random variables $n\in\{3,\ldots,9\}$ and the number of proof atoms $k$.
\State Generate $k$ Shannon-type atoms $S_1,\ldots,S_k$ from
$H(A)$, $H(A|B)$, $I(A;B)$, and $I(A;B|C)$.
\State Generate conditions by random sampling and LP feasibility checking, and split them into equality conditions
$\mathcal{C}_{=}$ and inequality conditions $\mathcal{C}_{\ge}$.
\State Represent each equality condition $C=0$ as two inequalities
$C\ge 0$ and $-C\ge 0$.
\State Assign coefficients $\alpha_i>0$ and $\beta_t>0$ to Shannon atoms and
inequality conditions, and assign coefficients $\eta_j\neq 0$ to equality conditions.
\State Construct the initial residual
\[
R \gets
\sum_i \alpha_i S_i
+
\sum_{C_j\in \mathcal{C}_{=}} \eta_j C_j
+
\sum_{D_t\in \mathcal{C}_{\ge}} \beta_t D_t .
\]
\State Let $\mathcal{V}$ be the multiset of all candidate proof  with their coefficients.
\State Initialize the dataset $\mathcal{D}\gets [\ ]$.
\While{$\mathcal{V}$ is not empty }
    \State Let $\mathcal{N}^{-}(R)=\{q:R_q<0\}$ be the negative coordinates of $R$.
    \State Select the atom $(\lambda,v)\in\mathcal{V}$ that maximizes
    \[
    -\textstyle\sum_{q\in \mathcal{N}^{-}(R)} \lambda v_q .
    \]
    \State Format the intput $x$ using the current residual target $R\ge 0$ and all conditions.
    \State Format the output $y$ using the selected proof step $\lambda v\ge 0$.
    \State Add the prompt $(x,y)$  to $\mathcal{D}$.
    \State Update $R\gets R-\lambda v$ and remove $(\lambda,v)$ from $\mathcal{V}$.
\EndWhile
\State \Return $\mathcal{D}$.
\end{algorithmic}
\end{algorithm}

The prompt is tokenized and bounded to a maximum context length of 4096 or
8192 tokens, depending on the dataset variant. Each dataset contains approximately
389 million tokens.

\subsection{Model Architecture and Fine-Tuning}

We fine-tune Qwen3 models (0.6B and 1.7B parameters)~\cite{qwen3} using Low-Rank Adaptation (LoRA)~\cite{hu2021lora} with rank $r=64$ and $\alpha=128$.
Training uses the AdamW 8-bit optimizer with learning rate $1.5 \times 10^{-5}$, cosine schedule with 200 warmup steps, micro-batch size 4, gradient accumulation over 8 steps (effective batch size 32), weight decay 0.01, and BF16 mixed precision.
Each model is trained for 2 epochs on two NVIDIA RTX 5880 Ada Generation GPUs.
All runs use seed 42 for reproducibility.

\subsection{Guided Beam-Search Tree Search} \label{Guided Beam-Search Tree Search}

At inference time, the fine-tuned model proposes candidate proof steps given the current residual $R_t$ and available conditions.
We employ a beam search over proof trees with configurable beam size $B$, rollouts per node $K$, and maximum depth $D$.
Objective is to find any path from $R_0$ to $R_t = 0$.

\textbf{Scoring and pruning.}
After generating $K$ candidate inequalities at a node, each valid candidate produces a new residual vector.
Nodes are scored by the greedy heuristic $\mathrm{score}(\mathbf{c}) = \sum_{i: c_i < 0} |c_i| \times 10^3$, which is the sum of absolute value of negative coefficients of the  residual inequality.
We prioritizes residuals with the smallest score, favor beam toward residuals that are closest to being non-negative in every coefficients---and thus closest to being provable.
Ablation with uniform random scoring (replacing this heuristic) reduces proof success from 83.3\% to 23.3\% (7/30), confirming the scoring heuristic is essential for effective beam allocation.

\textbf{Candidate validation pipeline.}
Each candidate step is validated through checks before being added to the beam:
\begin{enumerate}
  \item \textbf{Format filter}: Accepts only positive scalar multiples of froms
$H(A)$, $H(A|B)$, $I(A;B)$, and $I(A;B|C)$, or of known conditions. Multi-term Shannon sums not matching any condition are rejected.
  \item \textbf{Obviously-negative pruning}: Discards residuals where all coefficients are $\le 0$ ( residuals never reach zero).
  \item \textbf{LP false pruning} (applied on sample with $n \le 9$ only): When the residual and all conditions have $\le 9$ variable involved, we use a Shannon-type LP to detect impossible residuals and prune them. For $n \ge 10$, the LP can no longer incorporate all $n$ variables alongside the conditions; Thus, LP pruning is disabled. The ablation experiment shows that whether we enable of disable this function have minor impact on accuracy.
\end{enumerate}

\subsection{Datasets and Training Setup}

We construct four training datasets (A/B/C/D) varying two factors: maximum context length (4096 vs.\ 8192 tokens) and data distribution ($n=9$-skewed and not skewed), each containing 389.5M tokens:
Dataset A (508,803 rows, 4096, 6.0\% $n=9$);
Dataset B (256,870 rows, 4096, $n=9$-skewed, 52.6\% $n=9$);
Dataset C (8192, 8.0\% $n=9$, 77\% of rows exceed 3072 tokens);
Dataset D (8192, $n=9$-skewed, 66.9\% $n=9$, 95\% long-token share).

\subsection{Training Budget Selection}

We first train a 0.6B pilot run on Dataset A with 10 uniformly spaced checkpoints (fractions 0.1--1.0 of the full training schedule) and evaluate all checkpoints on a held-out validation set ($n=10$--$15$, 30 problems) to select a shared training-budget fraction.

\begin{table}[htbp]
  \renewcommand{\arraystretch}{1.2}
  \caption{Dataset A checkpoint sweep on the 30 problems.}
  \label{tab:budget_sweep}
  \centering
  \begin{tabular}{c|c|c|c}
    \toprule
    Fraction & Step & Pass Rate & Avg.\ Tree (Proved) \\
    \midrule
    0.2 & 3178  & 21/30 (70.0\%) & 514.0 \\
    0.4 & 6356  & 22/30 (73.3\%) & 478.7 \\
    \textbf{0.6} & \textbf{9534}  & \textbf{24/30 (80.0\%)} & \textbf{385.3} \\
    0.8 & 12712 & 23/30 (76.7\%) & 393.6 \\
    1.0 & 15885 & 25/30 (83.3\%) & 416.0 \\
    \bottomrule
  \end{tabular}
\end{table}

Table~\ref{tab:budget_sweep} shows the sweep results.
Performance plateaus around fraction 0.5--0.6, with checkpoint 9534 (fraction 0.6) achieving 80.0\% success and the lowest average proof tree size (385.3 nodes) among all high-performing fractions.
Later checkpoints show marginally improved success (83.3\% at fraction 1.0) but with reduced search efficiency (416.0 tree size), consistent with overfitting to training data patterns without improving underlying reasoning capability.
Fraction 0.6 is also the earliest checkpoint within a small margin of the best validation score, satisfying our preference for conservative early stopping.

We freeze fraction 0.6 as the shared training budget and apply it to all subsequent 0.6B runs (B/C/D) and the 1.7B run on Dataset A.

\section{Experimental Results}

\subsection{Evaluation Setup}

We evaluate on three test regimes:
\begin{itemize}
  \item \textbf{Easy set (E)}: 30 synthetic inequalities, $n=10$--$15$, five per $n$, with mixed conditions.
  \item \textbf{Hard set (H)}: 30 difficult inequalities, $n=8$--$10$, five per $n$. Half are short inequalities that our prototype model cannot solve, while the other half are provable only with large tree sizes ($>1000$ nodes). This set was designed to stress-test both model capability and search depth.
  \item \textbf{n=9 set (N9)}: An $n=9$ set, target plus condition token within 800--4096, generated by Alg. \ref{alg:mix-type-generation}.
  \item \textbf{Long tokens set (L)}: Token lengths $\geq$ 4096 for $n=7,8,10$, five per $n$, generated by Alg. \ref{alg:mix-type-generation}.
  \item \textbf{Final test set}: 60 inequalities, $n=10$--$15$, 10 per $n$, generated by Alg. \ref{alg:mix-type-generation}. We permuting $n=10$ samples index randomly to obtain $n=11$--$15$ sample to avoid large computing power required for direct LP generate.
  
\end{itemize}

Evaluation on N9 and L set is to ablate the effectiveness of each dataset on specific token-length and $n$-values portions.
Default search parameters are beam size 24, rollouts 4, maximum depth 20.
Evaluation uses 6--9 instances of LLMs parallely on an NVIDIA RTX 5880 Ada Generation cluster.

\subsection{ABCD Dataset Ablation (0.6B)}

Table~\ref{tab:scaling_abcd} reports the 0.6B model performance across all four datasets at the 0.6 training fraction.

\begin{table}[htbp]
  \renewcommand{\arraystretch}{1.2}
  \caption{Model size scaling and ablation across all regimes. E=Easy, H=Hard (30 each); N9=$n{=}9$ spec., L=Long (15 each).}
  \label{tab:scaling_abcd}
  \centering
  \scriptsize
  \begin{tabular}{c|c|c|c|c}
    \toprule
    Model / Dataset & E (30) & H (30) & N9 (15) & L (15) \\
    \midrule
    A (4096 Bal.) & 24/30 (80\%) & 9/30 (30\%) & \textbf{14/15 (93\%)} & \textbf{12/15 (80\%)} \\
    B (4096 Skew) & 24/30 (80\%) & 9/30 (30\%) & 13/15 (87\%) & \textbf{12/15 (80\%)} \\
    C (8192 Bal.) & 21/30 (70\%) & 9/30 (30\%) & \textbf{14/15 (93\%)} & 11/15 (73\%) \\
    D (8192 Skew) & \textbf{25/30 (83\%)} & 8/30 (27\%) & 13/15 (87\%) & 9/15 (60\%) \\
    \midrule
    1.7B A & 24/30 (80\%) & \textbf{14/30 (47\%)} & -- & -- \\
    \bottomrule
  \end{tabular}
\end{table}

On the easy set, all datasets perform comparably (70--83\%), with Dataset D marginally leading.
On the hard set, Datasets A/B/C tie at 30\% while D trails at 27\%.
Critically, while A, B, and C achieve the same hard-set pass rate, Dataset A attains this with a substantially smaller average tree size (1076.1 vs.\ 1257.8--1319.0), indicating more efficient proof paths and less susceptibility to search noise while accuracy is about the same as other dataset.
On specialist subsets, not skewed models (A/C) lead the $n=9$ set (93\% vs.\ 87\% for skewed), and 4096-token models (A/B) lead the long stress-test (80\% vs.\ 60\%--73\% for 8192-token models).
The interaction is clearest in Dataset D: it leads on the easy set but drops to last on the hard set and long stress-test, suggesting that skewed $n=9$ training at 8192 tokens encourages proof patterns that do not transfer to harder problems.

Based on these results, we select Dataset A (4096 not skewed) as the best overall configuration for its consistent performance and superior search efficiency.

\subsection{Tree Search Hyperparameter Sweep}

After selecting Dataset A, we conduct a systematic sweep over beam size $\{8, 16, 24, 32, 64\}$, rollouts $\{2, 4, 6, 8\}$, and maximum depth $\{16, 20, 24, 28\}$ on the hard set using the Dataset A model at fraction 0.6.

\begin{table}[htbp]
  \renewcommand{\arraystretch}{1.2}
  \caption{Selected tree search hyperparameter sweep results on the hard set (Dataset A, 0.6 fraction).}
  \label{tab:tree_sweep}
  \centering
  \begin{tabular}{c|c|c|c|c}
    \toprule
    Beam & Rollouts & Depth & Pass Rate & Avg.\ Tree \\
    \midrule
    24 & 4  & 20 & 9/30 (30.0\%)  & 1076.1 \\
    32 & 4  & 24 & 14/30 (46.7\%) & 2352.0 \\
    \textbf{32} & \textbf{6}  & \textbf{24} & \textbf{16/30 (53.3\%)} & \textbf{2588.7} \\
    32 & 8  & 24 & 12/30 (40.0\%) & 2117.0 \\
    64 & 6  & 24 & 10/30 (33.3\%) & 1524.8 \\
    \bottomrule
  \end{tabular}
\end{table}

The selected policy is beam size 32, rollouts 6, maximum depth 24, which achieves the highest success rate (53.3\% on the hard set).
Notably, performance degrades at the most aggressive settings (beam 64 or rollouts 8): log analysis confirms that failed high-parameter runs terminated naturally without hitting tree or time budgets, indicating that excessive breadth introduces distractor branches that steer the search away from valid proof paths.
This reveals a \emph{critical beam size} beyond which search quality declines.

\subsection{Scaling to 1.7B Parameters}

Table~\ref{tab:scaling_abcd} compares the 0.6B and 1.7B models, both trained on Dataset A at fraction 0.6 and evaluated under the identical search policy (beam 32, rollouts 6, depth 24).

The 1.7B model slightly underperformed compared to the 0.6B model on the test set (80.0\% vs.\ 83.3\%), the slightly 1.7B underperformed is by one sample, consistent with run-to-run stochastic variation rather than a systematic capability regression.
However, on the more difficult hard set ($n=8$--$10$, curated), the 1.7B model significantly outperforms the 0.6B model (46.7\% vs.\ 30.0\%).

\subsection{Final Test-Set Results and Comparison to LP-solver}

Table~\ref{tab:final_comparison_all}'s experiments on LP-solvers are in 300s time limit.(left) compares our LLM-based method with PSITIP ~\cite{psitip}. PSITIP can solve all the $n\leq 12$ samples but time out in bigger sample. Since LP-solver need to load elemental inequalities in $O(n^2 2^n)$ size. Our method do not need to load elemental inequalities, uses 2.33GB of VRAM at minimal setting. Since Aitip can only solve equality condition problems, we compare LP-solvers in such a dataset. Note that AITIP uses 36.5GB of RAM when solving $n= 15$ samples, indicating the memory limitation of the LP-solver method.

\begin{table}[htbp]
  \centering
  \caption{(Left)results dataset with inequality and equality conditions. (Right) LP-solvers benchmark, only equality conditions.}
  \label{tab:final_comparison_all}
  \scriptsize
  \begin{tabular}{c|c|c|c|c}
    \toprule
    & \multicolumn{2}{c}{\textbf{Testset}} & \multicolumn{2}{c}{\textbf{Equality Conditions}} \\
    \cmidrule(lr){2-3} \cmidrule(lr){4-5}
    $n$ & \textbf{0.6B A} & \textbf{PSITIP} & \textbf{AITIP} & \textbf{PSITIP} \\
    \midrule
    10 & 10/10 (100.0\%) & 10/10 (100\%) & 10/10 (100\%) & 10/10 (100\%) \\
    11 &  9/10 (90.0\%)  & 10/10 (100\%) & 10/10 (100\%) & 10/10 (100\%) \\
    12 &  7/10 (70.0\%)  &  0/10 (0\%)   & 10/10 (100\%) &  4/10 (40\%)  \\
    13 &  8/10 (80.0\%)  &  0/10 (0\%)   &  8/10 (80\%)  &  1/10 (10\%)  \\
    14 &  9/10 (90.0\%)  &  0/10 (0\%)   &  0/10 (0\%)   &  2/10 (20\%)  \\
    15 &  8/10 (80.0\%)  &  0/10 (0\%)   &  1/10 (10\%)  &  1/10 (10\%)  \\
    \midrule
    Overall & \textbf{51/60 (85.0\%)} & \textbf{20/60 (33.3\%)} & \textbf{39/60 (65.0\%)} & \textbf{28/60 (46.7\%)} \\
    \bottomrule
  \end{tabular}
\end{table}

\subsection{SOTA LLM Baselines}

We establish external baselines on the same test set to contextualize the SFT model performance.

\textbf{Zero-shot prompting.}
GPT-5.5\cite{gpt55} and Claude-4.6 Sonnet\cite{claude46sonnet} were evaluated via their websites, their output is not restricted to atomic step. Under zero-shot prompting, GPT-5.5 proved $1$ out of $60$ samples while while Claude-4.6 Sonnet solved none. The only successful case is Sample 7, whose proof uses the given condition and can be decomposed as $I=6C_1+S$, where $S$ is a non-negative combination of Shannon-type inequalities. 

\textbf{Tree-search with frontier LLMs.}
We also try on replacing our fine-tuned model with GPT-5.4-mini\cite{gpt54mini} or DeepSeek V3.2\cite{deepseekv32} the beam-search framework, using the API (temperature=1.2, top\_p=0.9, max\_tokens=512).
Test with $n=3$, 5 problems, were completed: DeepSeek V3.2 proved 1/5 (20.0\%) and GPT-5.4-mini proved 0/5 (0.0\%).
DeepSeek V3.2 and GPT-5.4-mini produced 45--300 format-invalid proposals per run (often multi-term expressions or natural-language fragments) and propose Shannon-infeasible output, showing that the format discipline required by the verifier is not satisfied by general-purpose LLMs.

\subsection{Search Score Function Ablation}
We tested on dataset E with random scoring on for beam selection. Accuracy drops from 83.3\% (25/30) to 23.3\% (7/30).
The average tree size on successful proofs are 1100.3 and 244.1 respectively for greedy function and uniform random scoring function. Only samples with small proof tree size are  proved, indicating  that uniform random scoring decrease tree search ability on proving samples that require large depth.

\section{Failure Analysis}

\subsection{Format Failure}

In the 0.6B Dataset A test-set evaluation, at $n=11$ and $n=13$, the model repeatedly proposes multi-term Shannon sums instead of atomic single-term expressions.
For example, on Sample 6 ($n=11$), the model proposes
$6\cdot(7H(X_{1,2,3,4,5,6}) - 5H(X_{1,4,6,7}) + \cdots) \ge 0$,
which is rejected by the format filter since it is not a positive scalar multiple of any provided condition. We suspect the model sometimes output complex pre-training-style expressions, conflicting with the strict atomic-step constraint.


\subsection{Step Quality Degradation}

On the failed $n=15$ samples in the final test-set evaluation, the model often produced format-valid but weak steps.
For instance, on a residual with 20+ terms and large coefficients (magnitudes 10--100), the model repeatedly proposes single-unit monotonicity steps (e.g., $H(X_{1,2}|X_{3}) \ge 0$) that reduce the residual by only a few terms with coefficient~1 while leaving large negative coefficients untouched.
Performance degrades at aggressive search settings (beam 64, rollouts 8): at high beam size and at high rollouts, it repeatedly generates near-duplicate atomic terms that only differ in coefficient (e.g., $H(X_{2,5}|X_{1,3,6,7,8}) \ge 0$ appears 7 consecutive times at rollout~8) rather than exploring diverse steps.
The search budget is consumed by these steps that make negligible progress toward zero.

\section{Conclusion}

We have shown that a 0.6B-parameter language model, fine-tuned on atomic entropy inequality proof steps and combined with guided beam search, can prove 85\% of test set  where GPT-5.5 scores only 1.7\% (1/60) under zero-shot prompting. 
A 1.7B model under identical conditions outperforms the 0.6B model on a hard set (46.7\% vs.\ 30.0\%), demonstrating that larger capacity aids deeper reasoning on complex problems at moderate $n$.
Systematic ablation reveals that a 4096-token not skewed training distribution yields the most robust performance, with no benefit from extended context, and that the beam-scoring heuristic is essential (random scoring reduces success rate from 83\% to 23\%).
Our method also solves more $n \geq 12$ problems compare to LP-solver under time limit.
The primary obstacles to further improvement are format failures at moderate $n$ and search budget exhaustion at high $n$, both of which suggest concrete directions for future work in constrained decoding, larger-scale search, and reinforcement learning~\cite{yue2025rl}.

Another way to speed up tree search is to strengthen LP false pruning.  When $n\ge 10$, we can randomly set some variables be constants or equal. A true information inequality must still be true after these changes. Hence, if the simplified inequality is false, then the original inequality cannot be true. This gives an inexpensive and efficient way to reject bad branches early, which may be investigated in future research. 

\section*{Acknowledgment}

The work of Shaocheng Liu and Cheuk Ting Li was partially supported by two grants from the Research Grants Council of the Hong Kong Special Administrative Region, China [Project No.s: CUHK 24205621 (ECS), CUHK 14209823 (GRF)].
The authors thank Chih Wei Ling, Yanxiao Liu and Yunhe Li for the insightful discussions.
\newpage

\end{document}